\documentclass[aps,epsfig,rotate,preprint]{revtex4}
\usepackage{amssymb}
\usepackage{latexsym}
\usepackage{amsfonts}
\usepackage{graphicx}
\usepackage{epsfig}
\usepackage{amsmath}
\usepackage{float}

\begin{document}

\title{Scaling and memory of intraday volatility return intervals in stock market}

\author{Fengzhong Wang$^1$, Kazuko Yamasaki$^{1,2}$, Shlomo Havlin$^{1,3}$ and H. Eugene~Stanley$^1$ }

\affiliation{$^1$Center for Polymer Studies and Department of Physics, Boston University, Boston, MA 02215 USA\\$^2$Department of Environmental Sciences, Tokyo University of Information Sciences, Chiba 265-8501,Japan\\$^3$Minerva Center and Department of Physics, Bar-Ilan University, Ramat-Gan 52900, Israel}

\date{9 November 2005 version ~~~ wyhs.tex}

\begin{abstract}

We study the return interval $\tau$ between price volatilities that are
above a certain threshold $q$ for 31 intraday datasets, including the
Standard \& Poor's 500 index and the 30 stocks that form the Dow Jones
Industrial index. For different threshold $q$, the probability density
function $P_q(\tau)$ scales with the mean interval $\bar{\tau}$ as
$P_q(\tau)={\bar{\tau}}^{-1}f(\tau/\bar{\tau})$, similar to that found
in daily volatilities. Since the intraday records have significantly 
more data points compared to the daily records, we could probe for much 
higher thresholds $q$ and still obtain good statistics. We find that the 
scaling function $f(x)$ is consistent for all 31 intraday datasets in 
various time resolutions, and the function is well approximated by the 
stretched exponential, $f(x)\sim e^{-a x^\gamma}$, with $\gamma=0.38\pm 0.05$ 
and $a=3.9\pm 0.5$, which indicates the existence of correlations. We 
analyze the conditional probability distribution $P_q(\tau|\tau_0)$ for 
$\tau$ following a certain interval $\tau_0$, and find $P_q(\tau|\tau_0)$ 
depends on $\tau_0$, which demonstrates memory in intraday return intervals. 
Also, we find that the mean conditional interval $\langle\tau|\tau_0\rangle$ 
increases with $\tau_0$, consistent with the memory found for 
$P_q(\tau|\tau_0)$. Moreover, we find that return interval records have long 
term correlations with correlation exponents similar to that of volatility 
records.

\end{abstract}


\maketitle

\section{introduction}

Statistical properties of price fluctuations
\cite{Kondor99,Lauritsen99,Mantegna91,Takayasu97,Mandelbrot63,%
Marsili98,Mantegna00,Vandewalle98,Sornette96,Bouchaud03,Micciche02,%
Picozzi02,Krawiecki02,Thomakos02,Lillo00} are very 
important to understand and model financial market dynamics, which 
has long been a focus of economic research. Stock volatility is of 
interest to traders because it quantifies risk, optimizes the portfolio
\cite{Bouchaud03,Yamasaki06,Johnson03} and provides a key input of option 
pricing models that are based on the estimation of the volatility of the 
asset \cite{Johnson03,Black73,Cox76,Cox79}. Although the logarithmic 
changes of stock price from time $t-1$ to time $t$, 
\begin{equation}
\label{e1}
G(t)\equiv\log\left({p_t\over p_{t-1}}\right) 
\end{equation}
is uncorrelated, their absolute values are known to be long-term power-law 
correlated \cite{Pagan96,Ding83,Dacorogna93,Wood85,Harris86,Admati88,%
Granger96,Bollerslev92,Cont98,Liu97,Cizeau97,Liu99}. The probability 
density function (pdf) of $G(t)$ possesses a power-law distribution 
\cite{Mandelbrot63}, 
\begin{equation}
\label{e2}
\Phi(G)\sim G^{-(\zeta+1)}, 
\end{equation}
with $\zeta\approx 3$ \cite{Liu99,Gopikrishnan99,Mantegna00,Gabaix03}. 
Also, $n_q(t)$, the number of times that the volatility $|G(t)|$ exceeds 
a threshold $q$, follows a power-law in the time $t$ after a market crash,
\begin{equation}
\label{e3}
n_q(t)\sim t^{-p}, 
\end{equation}
with $p\approx1$ \cite{Lillo03}. Eq.~(3) is the financial analog of the 
Omori earthquake law \cite{Omori94}. 

Recently Yamasaki et al. \cite{Yamasaki05} studied the behavior of
return intervals $\tau$ between volatilities that are above a certain
threshold $q$ [illustrated in Fig.~1(a)]. They analyzed {\it daily\/}
financial records and found scaling and memory in return intervals,
similar to that found in climate data \cite{Bunde05}. To investigate the
generality of these statistical features of Ref.~\cite{Yamasaki05}, here 
we examine 31 {\it intraday} datasets. We find that similar scaling and
memory behavior occurs at a wide range of time resolutions (not only on
the daily scale). Due to the larger size of the datasets we analyze, we 
are able to extend our work to significantly larger values of $q$. 
Remarkably, scaling functions are well approximated by the stretched 
exponential form, which indicates long range correlations in volatility 
records \cite{Bunde05}. Also, we explore clusters of short and long return 
intervals, and find that the larger is the cluster the stronger is the memory.

\section{databases analyzed}

We analyze the trades and quotes (TAQ) database from New York Stock
Exchange (NYSE), which records every trade for all the securities in
United States stock market for the two-year period from January 1, 2001
to December 31, 2002, a total of 497 trading days. We study all 30
companies of the Dow Jones Industrial Average index (DJIA). The 
sampling time is 1 minute and the average size is about
160,000 values per DJIA stock.  Another database we analyze is the
Standard and Poor's 500 index (S\&P 500), which consists of 500
companies.  This database is for a 13-year period, from January 1, 1984
to December 31, 1996, with one data point every 10 minutes (total data
points is about 130,000). For both databases, the records are
continuous in regular open hours for all trading days, due to the
removal of all market closure times.

\section{volatility definition}

In contrast to daily volatilities, the intraday data are known to show
specific patterns \cite{Liu99,Harris86,Admati88}, due to different
behaviors of traders at different periods during the trading day. For
example, the market is very active immediately after the opening
\cite{Admati88}, due to information arriving while the market is closed.
To understand the possible effect on volatility correlations, we
investigate the daily trend in DJIA stocks. The intraday pattern,
denoted as $A(s)$ \cite{Liu99}, is defined as
\begin{equation}
A(s)\equiv \frac{\sum_{i=1}^N |G^i(s)|}{N},
\end{equation}
which is the return at a specific moment $s$ of the day averaged over all
$N$ trading days, and $ G^i(s)$ is the price change at time $s$ in day
$i$. As shown in Fig. 1(b), the intraday pattern $A(s)$ has similar
behavior for the four stocks AT\&T, Citi, GE, IBM and the average over
30 DJIA stocks. The pattern is not uniformly distributed, exhibiting a
pronounced peak at the opening hours and a minimum around time $s=200$
min, that may cause some artificial correlations. To avoid the effect of
this daily oscillation, we remove the intraday pattern by studying
\begin{equation}
G^{\prime}(t)\equiv |G(t)|/A(s).
\end{equation}

In order to compare different stocks, we define the normalized
volatility $g(t)$ by dividing $G^{\prime}(t)$ with its standard
deviation,
\begin{equation}
g(t)\equiv \frac{G^{\prime}(t)}{(\langle G^{\prime}(t)^2\rangle-\langle G^{\prime}(t)\rangle^2)^{1/2}},
\end{equation}
where $\langle...\rangle$ is the time average for each separate
stock. Consequently, the threshold $q$ is measured in units of the
standard deviation of $G^{\prime}(t)$. As shown in Fig. 1(a), every
volatility $g(t)$ above a threshold $q$ (``event'') is picked and the
series of the time intervals between those events, $\{\tau(q)\}$, is
generated. The series depends on the threshold $q$. To maintain good
statistics and avoid spurious discreteness effects \cite{Yamasaki05}, 
we restrict ourselves to thresholds with average intervals
$\bar{\tau}=\bar{\tau}(q)>3$ time units (30 minutes for the S\&P 500 
and three minutes for the 30 stocks of the DJIA).

\section{scaling properties}

We study the dependence of $P_q(\tau)$ on $q$ , where $P_q(\tau)$ is the 
pdf of the return interval series $\{\tau(q)\}$. Figure~2 shows results 
for the S\&P~500 index and for two typical DJIA stocks, Citi and GE. The 
time window $\Delta t$ of volatility records is 1 minute for the DJIA 
stocks and 10 minutes for the S\&P 500. The left panels of Fig.~2 
[(a), (c), (e)], show that the pdf $P_q(\tau)$ for large $q$ decays slower
than for small $q$. The right panels of Fig.~2 [(b), (d), (f)]
show the {\it scaled\/} pdf $P_q(\tau)\bar{\tau}$ as a function of
the {\it scaled\/} return intervals $\tau/\bar{\tau}$.
The five curves for $q=2$, 3, 4, 5 and 6 collapse onto a
single curve. Thus the distribution functions follow the scaling
relation \cite{Yamasaki05,Bunde94}
\begin{equation}
P_q(\tau)=\frac{1}{\bar{\tau}}f(\tau/\bar{\tau}).
\end{equation}
We also study the other 28 DJIA stocks and find that they have similar 
scaling behavior for different thresholds. 

To examine the scaling for larger thresholds with good statistics, we 
calculate the return intervals of each DJIA stock, and then aggregate 
all the data. As shown in Fig.~2(g) and (h), the scaling behavior extends 
even to $q=15$. In Eq.~(7), the scaling function $f(\tau/\bar{\tau})$ 
does not directly depend on the threshold $q$ but only through 
$\bar{\tau}\equiv\bar{\tau}(q)$. Therefore, if $P_q(\tau)$ for an 
individual value of $q$ is known, distributions for other thresholds 
can be predicted by the scaling Eq.~(7). In particular, the distribution 
of rare events (very large $q$, such as market crashes) may be extrapolated 
from smaller thresholds, which have enough data to achieve good statistics.

Next, we investigate the similarity of scaling functions for different
companies. Scaled pdfs $P_q(\tau)\bar{\tau}$ with $q=2$ for return
intervals (upper symbols) are plotted in Fig. 3(a), showing the S\&P 500 
index and 30 DJIA stocks in alphabetical order of names (one symbol 
represents one dataset). We see that the pdf curves collapse, so their
scaling functions $f(x)$ are similar, consistent with a universal
structure for $P_q(\tau)$. As suggested by the line on upper symbols in
Fig. 3(a) and on the filled symbols in Fig. 4, the function $f(x)$ may
follow a stretched exponential form \cite{Bunde05}, 
\begin{equation}
f(x)\sim e^{-ax^\gamma}.
\end{equation}
Remarkably, we find that all 31 datasets have similar exponent values, and
conclude that $\gamma$ appears to be ``universal'', with 
\begin{equation}
\gamma=0.38\pm 0.05.
\end{equation}
The value $a$ is found to be almost the same for all datasets, 
\begin{equation}
a=3.9\pm 0.5.
\end{equation}
Further, we plot the stretched exponential fit for four companies, 
AT\&T, Citi, GE and IBM in a log-linear plot [Fig.~3(b)]. We find 
good fits for all four companies, and we also find good collapse for 
different thresholds for each stock. The scaling function clearly 
differs from the Poisson distribution for uncorrelated data, 
$f(x)\sim e^{-x}$, which is demonstrated by curves on lower symbols 
in Fig.~3(a).

For statistical systems, the time resolution of records is an important
aspect. The system may exhibit diverse behavior in different time
windows $\Delta t$. In Fig.~4 we analyze five time scales for four typical 
companies ($q=2$): (a) AT\&T, (b) Citi, (c) GE and (d) IBM. It is seen that
for $\Delta t=1$, 5, 10, 15, and 30 minutes, the $P_q(\tau)\bar{\tau}$ curves
collapse onto one curve, which shows the persistence of the scaling for
a broad range of time scales. Thus there seems to be universal structure
for stocks not only in different companies, but also in each stock with
various time resolutions.

To understand the origin of the scaling behavior in return intervals, 
we analyze pdfs of the volatility after shuffling (in order to remove
correlations in the volatility records \cite{Liu99,Yamasaki05}). For
uncorrelated data, as expected, a Poisson distribution is obtained,
shown by the lower symbols in Fig.~3(a) and empty symbols in Fig.~4. 
In contrast to the distribution for uncorrelated records, the distribution
 of the actual return intervals (the upper symbols in Fig.~3(a) and 
filled symbols in Fig.~4) is more frequent for both small and large
intervals, and less frequent in intermediate intervals. The distinct
difference between the distributions of return intervals in the original
data and shuffled records suggests that the scaling behavior and the form 
in Eq.~(8) must arise from long-term correlations in the volatility (see 
also \cite{Bunde05}).

\section{Memory Effects}

The sequence of return intervals may, or may not, be fully characterized 
by $P_q(\tau)$, depending on the time organization of the sequence. If 
the sequence of return intervals are {\it uncorrelated}, they are independent 
of each other and totally determined by the probability distribution. On 
the other hand, if the intervals are {\it correlated}, the memory will also 
affect the order in the sequence of intervals. 

To investigate  the memory in the records, we study the conditional pdf, 
$P_q(\tau|\tau_0)$, which is the probability of finding a return interval 
$\tau$ immediately after a return interval of size $\tau_0$. In records 
without memory, $P_q(\tau|\tau_0)$ should be identical to $P_q(\tau)$ and 
independent of $\tau_0$. Otherwise, it should depend on $\tau_0$. Due to 
the poor statistics for a single $\tau_0$, we study $P_q(\tau|\tau_0)$ 
for a bin (range) of $\tau_0$. The entire database is partitioned into 
8 equal-size bins with intervals in increasing length. Fig.~5 shows
$P_q(\tau|\tau_0)$ for $\tau_0$ in the smallest (solid symbols) and
largest (open symbols) subset of the four stocks AT\&T, Citi, GE and 
IBM. For $\tau_0$ in the lowest bin the probability is larger for
small $\tau$, while for $\tau_0$ in the largest bin the probability is
higher for large $\tau$. Thus, large $\tau_0$ tend to be followed by
large $\tau$, while small $\tau_0$ tend to be followed by small
$\tau$ (``clustering''), which indicates memory in the return interval 
sequence. Thus, long-term correlations in the volatility records affect 
the pdf of intervals as well as the time organization of $\tau$. Note also 
that $P_q(\tau|\tau_0)$ for all thresholds seems to collapse onto a single
scaling function for each of the $\tau_0$ subsets. 

Further, the memory is also seen in the mean conditional return 
interval $\langle\tau|\tau_0\rangle$, which is the first moment of
$P_q(\tau|\tau_0)$, immediately after a given $\tau_0$ subset.  Filled
symbols in Fig.~6 show again that large $\tau$ tend to follow large
$\tau_0$, and small $\tau$ follow small $\tau_0$, similar to the 
clustering in the conditional pdf $P_q(\tau|\tau_0)$. Correspondingly,
shuffled data (empty symbols) exhibits a flat shape, demonstrating that
the value of $\tau$ is independent on the previous interval $\tau_0$.

The quantities $P_q(\tau|\tau_0)$ and $\langle\tau|\tau_0\rangle$ show
memory for intervals that immediately follow an interval $\tau_0$, which
indicates short-term memory in the return interval records. To study the
possibility that the long-term memory exists in the return intervals 
sequence, we investigate the mean return interval after a cluster of 
$n$ intervals, all within a bin $\tau_0$. To obtain good statistics we 
divide the sequence only into two bins, separated by the median of the 
entire database. We denote intervals that are above the median by ``+'', 
and that are below the median by ``--''. Accordingly, $n$ consecutive 
``+'' or ``--'' intervals form a cluster and the mean of the return 
intervals after such $n$-clusters may reveal the range of memory in the 
sequence. Fig.~7 shows the mean return intervals 
$\langle\tau|\tau_0\rangle/\bar{\tau}$ vs. the size $n$, where $\tau_0$ 
in $\langle\tau|\tau_0\rangle/\bar{\tau}$ refers to a cluster with size 
$n$. For ``+'' clusters, the mean intervals increase with the size of the 
cluster, the opposite of that for ``--'' clusters. The results indicate 
long-term memory in the sequence of $\tau$ since we do not see a plateau 
for large clusters.

To further test the range of long-term correlations in the return
interval time series, we apply the detrended fluctuation analysis (DFA)
method \cite{Peng94,Hu01,Bunde00}. After removing trends, the DFA method
computes the root-mean-square fluctuation $F(\ell)$ of a time series
within windows of $\ell$ points, and determines the correlation exponent
$\alpha$ from the scaling function $F(\ell)\sim \ell^\alpha$. The
exponent $\alpha$ is related to the autocorrelation function exponent
$\gamma$ by 
\begin{equation}
\alpha=1-\gamma/2,
\end{equation}
and autocorrelation function $C(t)\sim t^{-\gamma}$ where $0<\gamma<1$ 
\cite{Bunde98,Bunde05}.  When $\alpha>0.5$, the time series has long-term 
correlations and exhibits persistent behavior, meaning that large values are 
more likely to be followed by large values and small values by small ones. 
The value $\alpha=0.5$ indicates that the signal is uncorrelated (white 
noise). 

We analyze the volatility series and the return interval series by using 
DFA method. The results of S\&P 500 index and 30 DJIA stocks for two regimes 
(split by $\ell^\ast=390$ for volatilities and $\ell^\ast=93$ for return 
intervals, which corresponds to 1 day in time scale) are shown in Fig.~8 
\cite{Peng94}. We see that $\alpha$ values are distinctly different in the 
two regimes, and both of them are larger than $0.5$, which indicates 
long-term correlations in the investigated time series but they are not the 
same for different time scales. For large scales ($\ell>\ell^\ast$), 
$\alpha=0.98\pm 0.04$ for the volatility (group mean$\pm$ standard deviation) 
and $\alpha=0.92\pm 0.04$ for the return interval are almost the same, and 
the differences are within the error bars. These results are consistent with 
Refs.~\cite{Liu99,Yamasaki05} for $\alpha$ of the volatilities, and with 
Ref.~\cite{Yamasaki05} for $\alpha$ of the return intervals. For short 
scales ($\ell<\ell^\ast$), we find $\alpha=0.66\pm0.01$ for the volatility 
(consistent with Ref.~\cite{Liu99}) and $\alpha=0.64\pm0.02$ of the return 
intervals, and the differences are again in the range of the error bars. 
Here error bars refer to that of each dataset, not the standard 
deviation of $\alpha$ group for $31$ datasets, and average error bars 
$\simeq 0.06$. Similar crossover from short scales to large scales with 
similar values of $\alpha$ have been also observed for intertrade times by
Ivanov et al. \cite{Ivanov04}. Such behavior suggests a common origin for
the strong persistence of correlations in both volatility and return
interval records, and in fact the clustering in return intervals is related 
to the known effect of volatility clustering \cite{Lux00,Giardina01,Lux02}.

\section{Discussion and conclusion}

The value of $\gamma\simeq 0.4$ could be a result of $\gamma=2-2\alpha$ 
from Eq.~(11), where $\alpha\simeq 0.8$ is the average of the two $\alpha$ 
regimes that we observe (see Fig.~8). It is possible for the value of 
$\gamma$ to be different for small and large $q$ values. The reason for 
this differences is that for small $q$ the low volatilities are probed
 and therefore the time scales are controlled by $\alpha\simeq 0.65$ 
(below the crossover), while for the large $q$ the high volatilities are 
probed, which represent large time scales (above the crossover), controlled 
by $\alpha\simeq 0.95$. We will undertake further analysis to test this 
possibility.

In summary, we studied scaling and memory effects in volatility return
intervals for {\it intraday\/} data. We found that the distribution function
for the return intervals can be well described by a single scaling
function that depends only on the ratio of $\tau/\bar{\tau}$ for DJIA
stocks and S\&P 500 index, for various time scales ranging from short term
$\Delta t=1$ minute to $\Delta t=30$ minutes. The scaling function,
which results from the long-term correlations in the volatility records,
differs from the Poisson distribution for uncorrelated data. We found
that the scaling function can be well approximated by the stretched
exponential form, $f(x)\sim e^{-a x^\gamma}$ with $\gamma=0.38\pm 0.05$ 
and $a=3.9\pm 0.5$.  We showed strong memory effects by analyzing the 
conditional pdf $P_q(\tau|\tau_0)$ and mean return interval 
$\langle\tau|\tau_0\rangle$. Furthermore, we studied the mean interval 
after a cluster of intervals, and found long-term memory for both clusters 
of short and long return intervals. We demonstrated by the DFA method that 
the volatility and return intervals have long-term correlations with 
similar correlation exponents.


\section*{Acknowledgments}

We thank L. Muchnik, J. Nagler and I. Vodenska for helpful discussions
and suggestions, and NSF for support.




\newpage

\begin{figure*}
\begin{center}
   \includegraphics[width=0.65\textwidth, angle = -90]{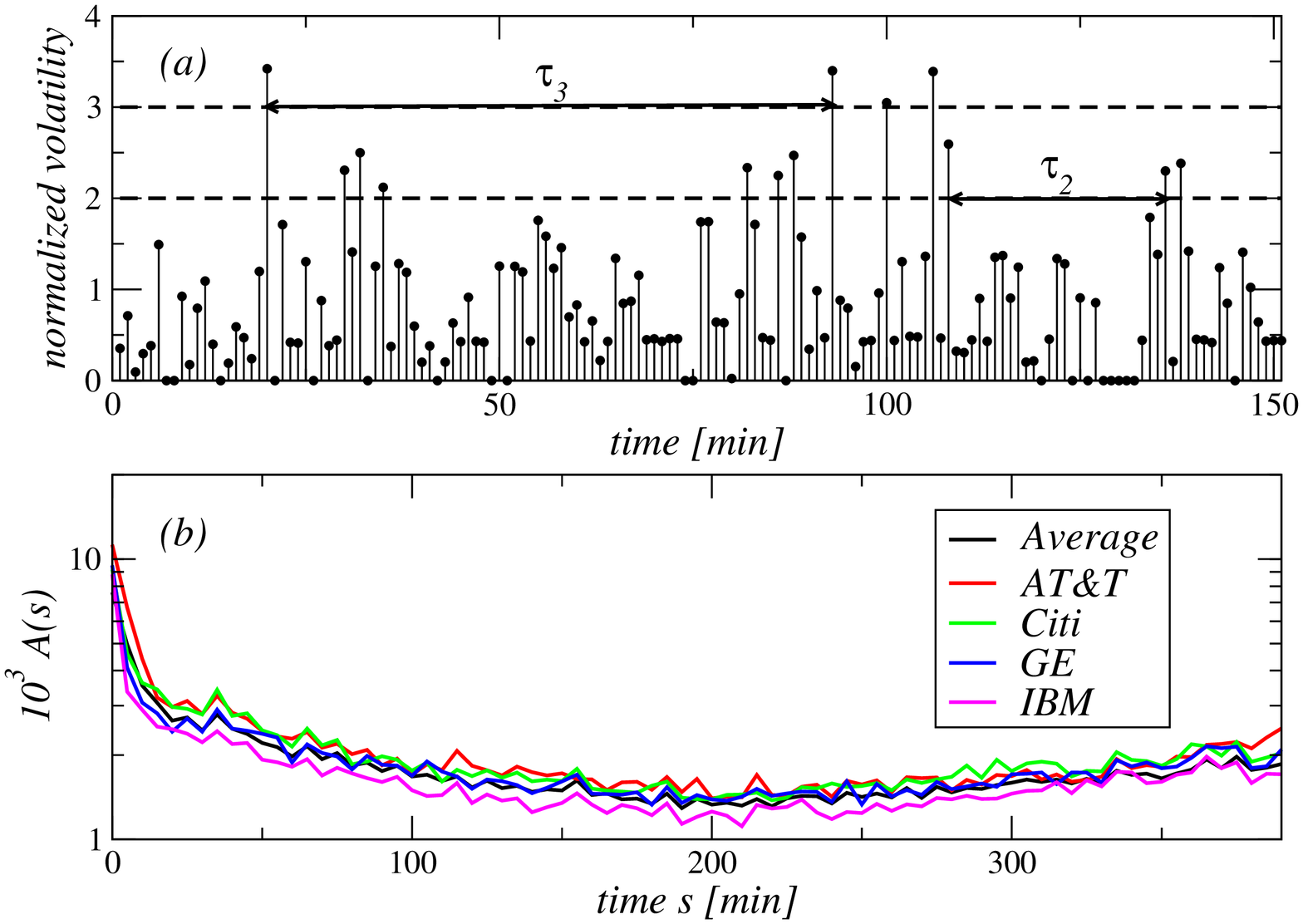}
\end{center}
\caption{(a) Illustration of volatility return intervals for a 
volatility time series for IBM on May 10, 2002. Return intervals 
$\tau_3$ and $\tau_2$ for two thresholds $q=3$ and $2$ are displayed. 
(b) The 5-min interval intraday pattern for AT\&T, Citi, GE, IBM
   and the average over 30 DJIA stocks. The time $s$ is the moment
  in each day, while $A(s)$ is the mean return over all trading
  days. Note that all curves have a similar pattern, such as a
  pronounced peak after the market opens and a minimum around noon ($s\simeq
  200$ min).}
\label{Fig1}
\end{figure*}

\begin{figure*}
\begin{center}
   \includegraphics[width=0.65\textwidth, angle = -90]{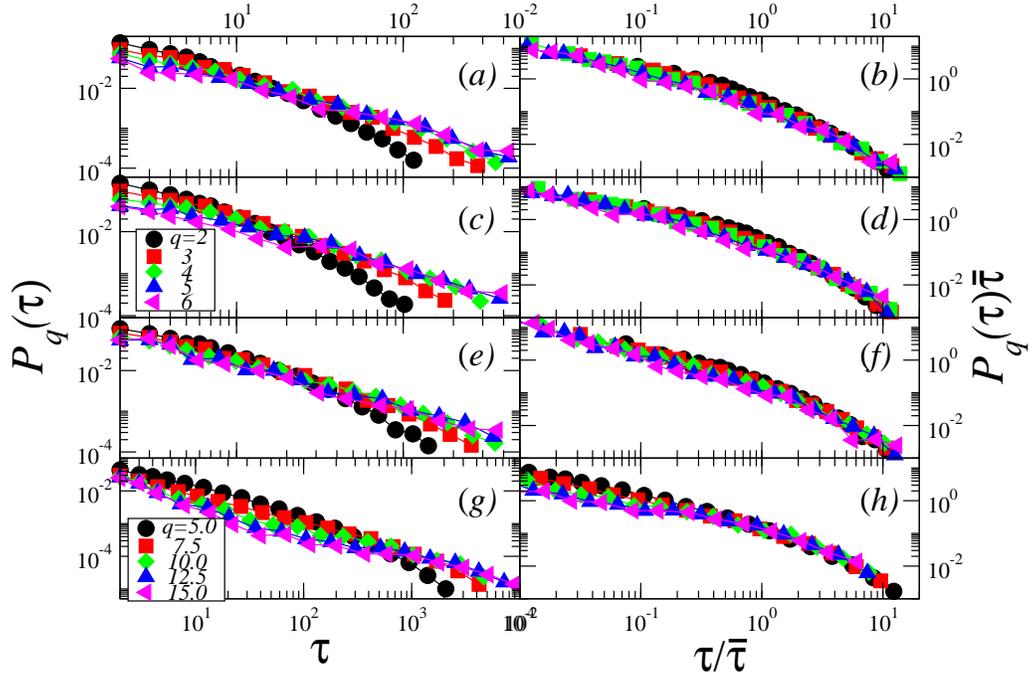}
\end{center}
\caption{Distribution and scaling of return intervals for (a \& b) Citi,
(c \& d) GE, (e \& f) S\&P 500 and (g \& h) mixture of 30 DJIA stocks (for
very large thresholds). Symbols are for different threshold $q$, as shown 
in (c) for (a) to (f) and shown in (g) for (g) and (h). The sampling time 
for S\&P~500 is 10 minutes, and for the stocks is 1 minute. For one dataset,
the distributions $P_q(\tau)$ are different with different $q$, but they
collapse onto a single curve for $P_q(\tau)\bar{\tau}$ vs. 
$\tau/\bar{\tau}$ ($\bar{\tau}$ is the mean interval), which indicates a 
scaling relation. (g) and (h) show that the scaling can extend to very 
large thresholds.}
\label{Fig2}
\end{figure*}

\begin{figure*}
\begin{center}
   \includegraphics[width=0.65\textwidth, angle = -90]{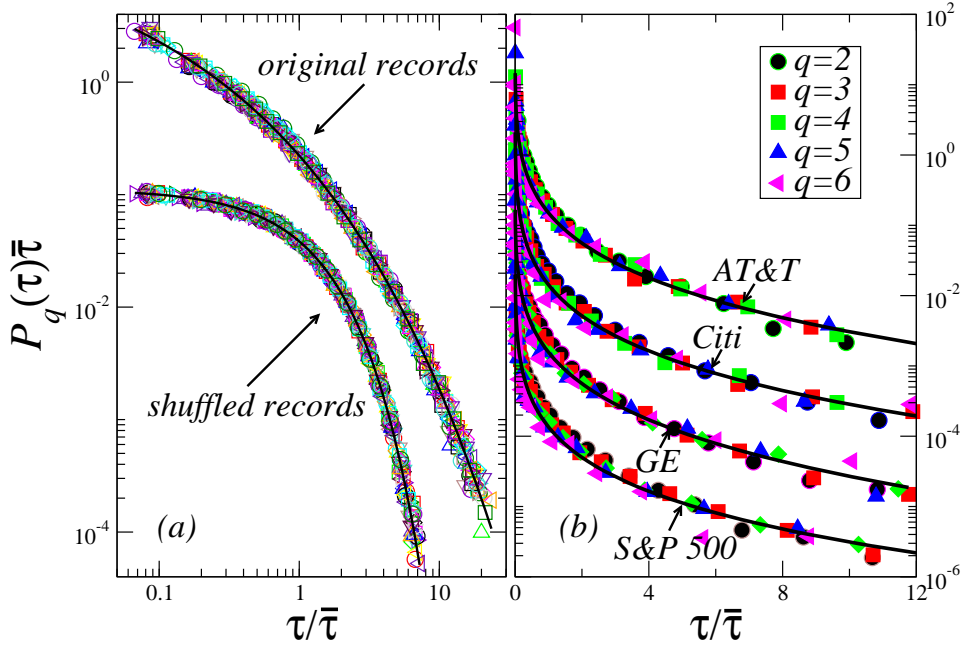}
\end{center}
\caption{(a) Scaling of return intervals for all 30 DJIA stocks and S\&P
500 index. Scaled distribution function $P_q(\tau)\bar{\tau}$
vs. $\tau/\bar{\tau}$ with threshold $q=2$ for actual return intervals, as 
well as for the shuffled volatility records (divided by 10) are shown. 
Every symbol represents one stock. The line on the symbols for original 
records suggests a stretched exponential relation, $f(x)\sim e^{-a x^\gamma}$ 
with $\gamma\simeq 0.38\pm 0.05$ and $a\simeq 3.9\pm 0.5$, while the curve 
fitting the shuffled records is exponential, $y=e^{-bx}$, from a Poisson 
distribution. Note that all the datasets are consistent with a single 
scaling relation. A Poisson distribution indicates no correlation in 
shuffled volatility data, but the stretched exponential behavior indicates 
strong correlation in the volatilities (see \cite{Bunde05}). (b) Stretched 
exponential fit for AT\&T, Citi, GE and S\&P 500 all with $\gamma\simeq 0.4$. 
Each stock is well approximated by stretched exponential for diverse 
thresholds, $q=2$ 3, 4, 5 and 6, presented in the plot. Each plot is shifted 
by $\times 10$ for clarity.}
\label{Fig3}
\end{figure*}

\begin{figure*}
\begin{center}
   \includegraphics[width=0.65\textwidth, angle = -90]{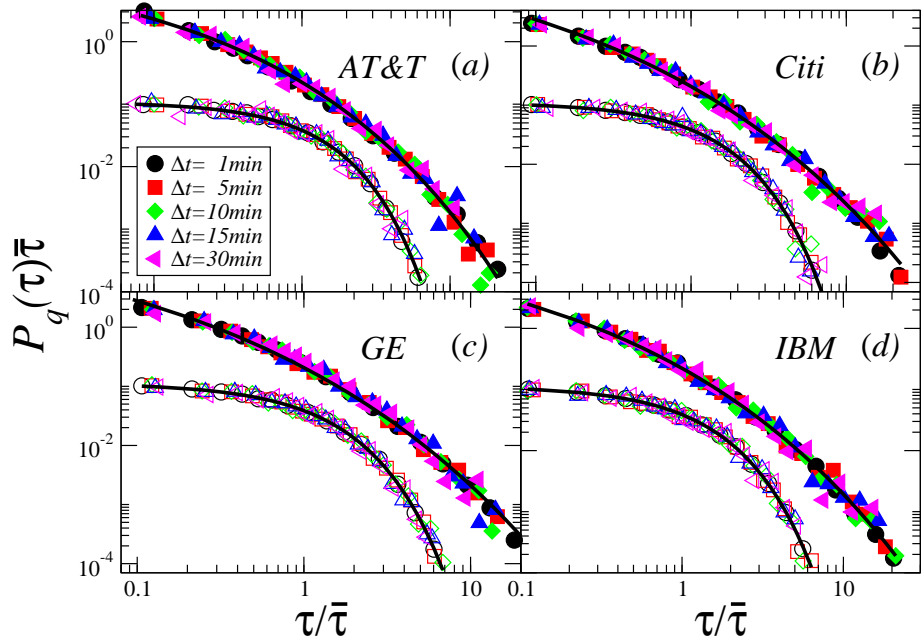}
\end{center}
\caption{Scaling for different time windows, $\Delta t=1,5,10,15$ and
$30$ min. Plots display scaled pdf $P_q(\tau)\bar{\tau}$ with threshold
$q=2$ for volatility return intervals (filled symbols) and shuffled
volatility records (shifted by factor 10, open symbols) vs. $\tau/\bar{\tau}$
of (a) AT\&T, (b) Citi, (c) GE and (d) IBM. Each symbol represents one
scale $\Delta t$, as shown in (a). Similar to Fig. 2 and Fig. 3, curves
fall onto a single line for actual return intervals and shuffled data
respectively, which indicates the scaling relation in Eq.~(6). Also, the 
actual return intervals suggest a stretched exponential scaling function,
demonstrated by the line fitting the solid symbols. The stretched exponential 
is a result of the long-term correlations in the volatility records. The
shuffled volatility records display no correlation, indicated by the
good fit (solid line) to the Poisson distribution.}
\label{Fig4}
\end{figure*}

\begin{figure*}
\begin{center}
   \includegraphics[width=0.65\textwidth, angle = -90]{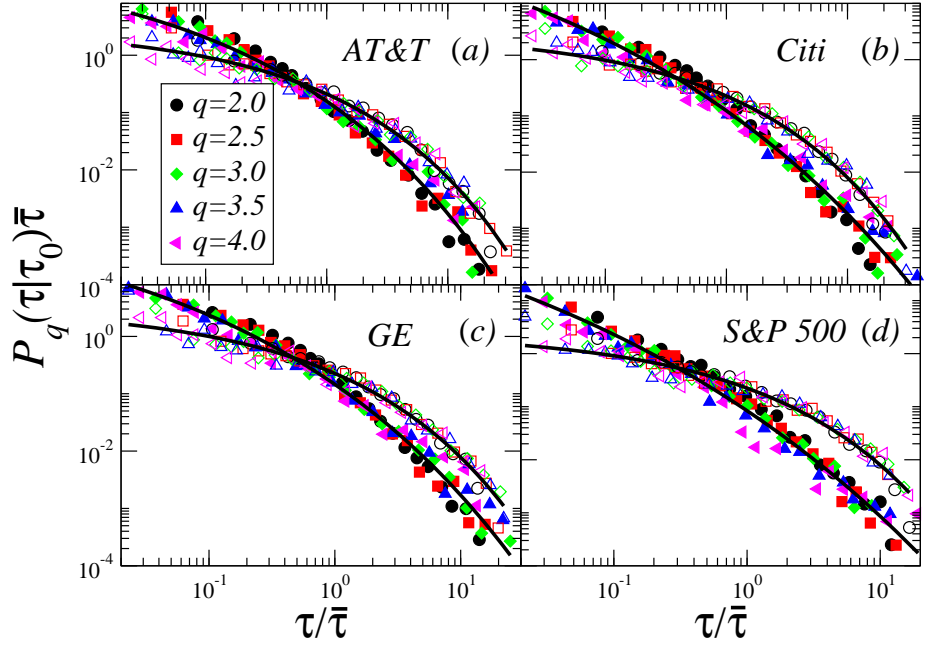}
\end{center}
\caption{Scaled conditional distribution $P_q(\tau|\tau_0)\bar{\tau}$
vs. $\tau/\bar{\tau}$ for (a) AT\&T, (b) Citi, (c) GE and (d) S\&P 500.
 Here $\tau_0$ represents binning of a subset which contains 1/8 of the
total number of return intervals in increasing order. Lowest 1/8 subset
(solid symbols) and largest 1/8 subset (empty symbols) are displayed,
which have different tendency, as suggested by black curves. Symbols are
plotted for different thresholds, denoted in (a). In contrast to the
largest subset, the lowest bin has larger probability for small intervals 
and smaller probability for large values, which indicates memory in 
records: small intervals tend to follow small ones and large intervals 
tend to follow large ones. Solid curves on symbols are stretched exponential 
fits.}
\label{Fig5}
\end{figure*}

\begin{figure*}
\begin{center}
   \includegraphics[width=0.65\textwidth, angle = -90]{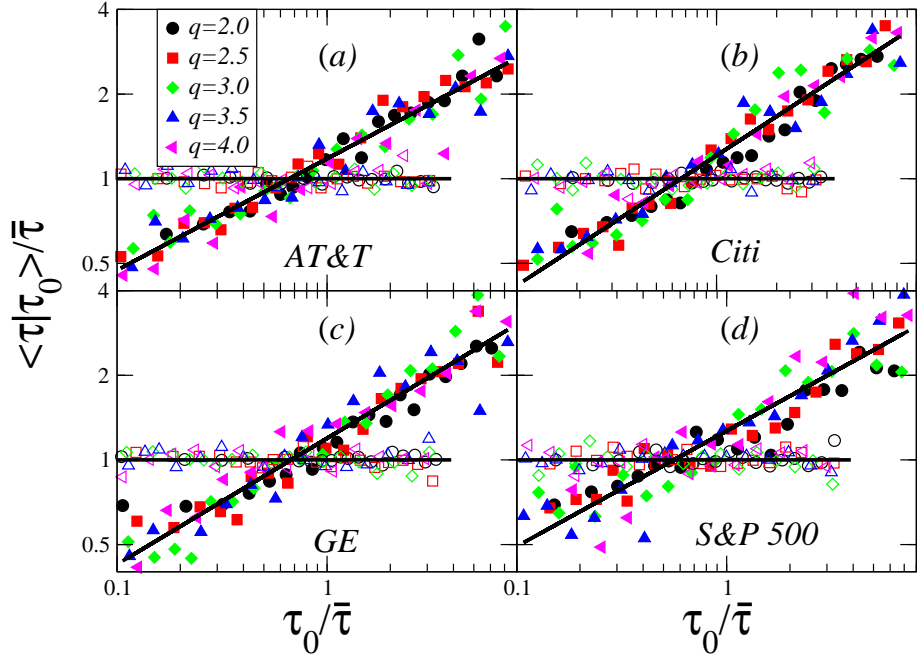}
\end{center}
\caption{Scaled mean conditional return interval
$\langle\tau|\tau_0\rangle/\bar{\tau}$ vs. $\tau_0/\bar{\tau}$ for 
(a) AT\&T, (b) Citi, (c) GE and (d) S\&P 500. The 
$\langle\tau|\tau_0\rangle/\bar{\tau}$ of intervals (filled symbols) 
and shuffled records (open symbols) are plotted. Five thresholds, 
$q=2.0$, 2.5, 3.0, 3.5 and 4.0 are represented by different symbols, 
as shown in (a). The distinct difference between actual intervals and 
shuffled records implies memory in the original interval records.}
\label{Fig6}
\end{figure*}

\begin{figure*}
\begin{center}
   \includegraphics[width=0.65\textwidth, angle = -90]{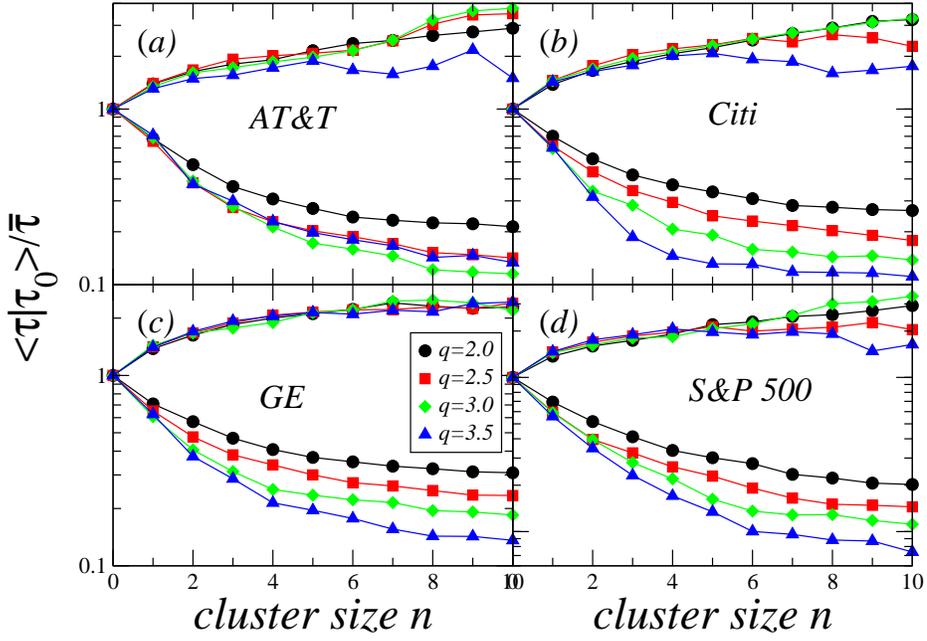}
\end{center}
\caption{Memory in return interval clusters. $\tau_0$ represents a
cluster of intervals, consisting of $n$ consecutive values that all are
above (denoted as ``+'') or below (``--'') the median of the entire
interval records. Plots display the scaled mean interval conditioned on 
a cluster, $\langle\tau|\tau_0\rangle/\bar{\tau}$, vs. the size $n$ of 
the cluster for (a) AT\&T, (b) Citi, (c) GE and (d) S\&P 500. One symbol
shows one threshold $q$, as shown in (c). The upper part of curves is
for ``+'' clusters while the lower part is for ``--'' clusters. The
plots show that ``+'' clusters are likely to be followed by large
intervals, and ``--'' clusters by small intervals, consistent with
long-term memory in return interval records.}
\label{Fig7}
\end{figure*}

\begin{figure*}
\begin{center}
   \includegraphics[width=0.65\textwidth, angle = -90]{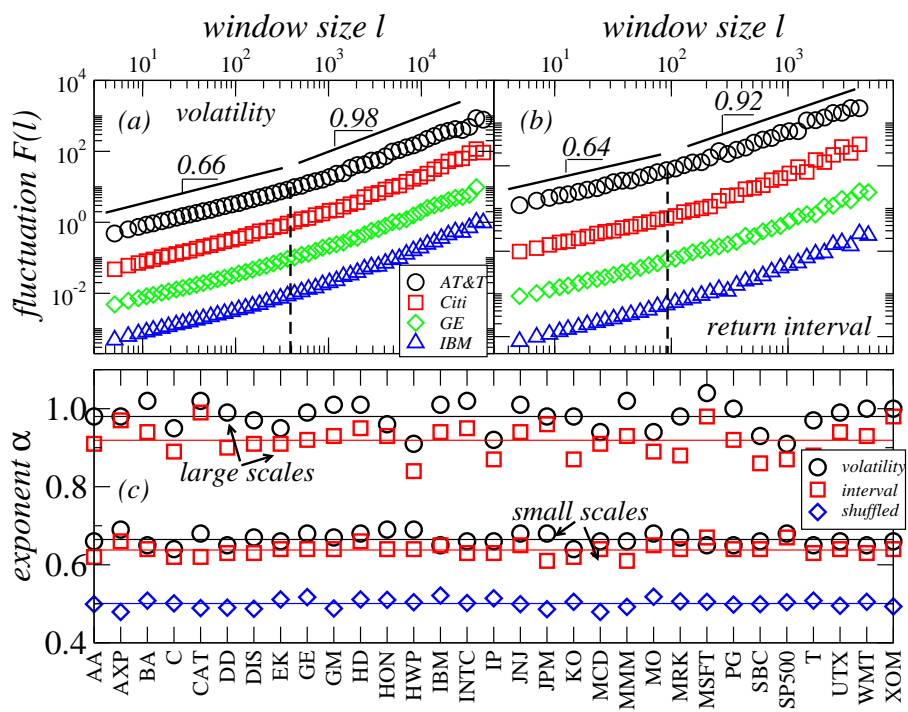}
\end{center}
\caption{Root mean square fluctuation $F(\ell)$ for (a) volatility 
records and (b) return interval records ($q=2$) obtained by the DFA 
method. Four companies are shown, AT\&T, Citi, GE and IBM (each shifted 
by factor of 10). The range of window size is split by vertical dashed 
lines, $\ell^\ast=390$ for volatilities (sampled each minute) and 
$\ell^\ast=93$ for return intervals, both corresponding to a time window 
of one day. The two regimes have different correlation exponents, as 
indicated by the straight lines. (c) Correlation exponent $\alpha$ for 
30 DJIA stocks and S\&P 500 index (related stock names are shown in 
x-axis). Volatility (circles) and return interval (squares) of large and 
smaller scales are shown. Note that most companies have smaller exponent 
for intervals than for volatilities, but their differences still are in 
the range of the error bars. Shuffled records (diamonds) possess $\alpha$ 
values around $0.5$ that indicate no correlation. Large scales 
($\alpha=0.98\pm0.04$ and $\alpha=0.92\pm0.04$, group average$\pm$standard 
deviation for volatilities and intervals respectively) and small scales 
($\alpha=0.66\pm0.01$ and $\alpha=0.64\pm0.02$ correspondingly) show 
different correlations for different scales, since $\alpha>0.5$.}
\label{Fig8}
\end{figure*}


\end{document}